# A Proposal for the Characterization of Multi-Dimensional Inter-relationships of RDF Graphs Based on Set Theoretic Approach

Ayan Chakraborty[1;4], Shiladitya Munshi[2;4], and Debajyoti Mukhopadhyay[3;4]

[1] Dept. of Computer Science Engineering
Techno India College of Technology, Kolkata- 700156, India,
achakraborty.tict@gmail.com
[2] Dept. of Information Technology
Meghnad Saha Institute of Technology,Kolkata- 700150,India
shiladitya.munshi@yahoo.com
[3] Department of Information Technology
Maharastra Institute of Technology,Pune: 411038, India
debajyoti.mukhopadhyay@gmail.com
[4] Web Intelligence & Distributed Computing Research Lab, Golf Green, Kolkata: 700095, India

Abstract. In this paper a Set Theoretic approach has been reported for analyzing inter-relationship between any numbers of RDF Graphs. An RDF Graph represents triples in Resource Description Format (RDF) of semantic web. So the identification and characterization of criteria for inter-relationship of RDF Graphs shows a new road in semantic search. Using set theoretic approach, a sound framing criteria can be designed that examine whether two RDF Graphs are related and if yes, how these relationships could be described with formal set theory. Along with this, by introducing RDF Schema, the inter-relationship status is refined into n-dimensional induced relationships.

## 1   Introduction

The Resource Description Framework (RDF) is a family of World Wide Web Consortium (W3C) specifications originally designed to express exchange and re-use structured metadata in semantic web. Using this simple model, it allows structured and semi-structured data to be mixed, exposed, and shared across different applications.

The RDF working group has chosen the right course in developing a simple and strictly-defined textual format for RDF graphs. This format is named N-Triples, and is incorporated into the RDF Test Cases working draft. N-Triples are line-oriented format. Each triple must be written on a separate line, and consists of a subject specifier, a predicate specifier, then an object specifier, followed by a period. One or more spaces or tabs separate subject from predicate, and predicate from object.

In this paper a Set Theory based approach is presented for identification as well as characterization of the relationship between two RDF Graphs.

## 2  Inter-relationship between RDF Graphs

In this paper RDF Graphs are represented with set diagrams. The intersection of given two RDFSets will denote the relationship between them. The set representation of an RDF consists of three subsets: subject, predicate and object. But existence of a blank node is critical to this point.

Let V be a vocabulary be an RDF Graph with vocab(T) V and $G_{dir;label;multi}$ the set of directed, edge and node-labeled multigraphs. We then define a map

: RDF Graph(V) ←$G_{dir;label;multi}$

as follows: (T)= (N,E,$L_N$ ,$L_E$), where

N={$n_x$:x∈Subj(T) ∪Obj(T)}
$L_N$ ($n_x$)= x,$d_x$      if x is literal and $d_x$ is datatype identifier
x          else

E={$e_{s,p,o}$:(S,P,O) ∈T}
from ($e_{s,p,o}$)=$n_s$,     to ($e_{s,p,o}$)=$n_o$     and $L_E$($e_{s,p,o}$)=P

The inter relationship between two RDF Graph is important from semantic search perspective. As mentioned in previous discussions, the semantic expressiveness of a statement can potentially be stored through RDF Graph and hence discovery of inter relationship criteria for RDF Graphs could form the basis of a formalized graph based search algorithm in the context of reservation and exploration of semantic nature of the statements or assertions.

On this background following section characterizes the criteria for establishing relationship between two RDF Graphs.

## 3  Characterization of inter-relationship between two RDF Graphs

The need of characterization of RDF Graph relationship is steeply growing as more and more data are being published in Semantic Web with RDF standard. The challenge of mining data from Semantic Web mostly can be met with the proper identification of related triples based on semantic constructs like subjects or objects etc. The network of those related RDF Graphs forms the local reference frame for the information to be searched.

On the context of present discussion following subsections characterizes RDF Graphs through a Set Theory approach.

### 3.1 Subject-Subject and Predicate-Predicate relationship

Subject-Subject and Predicate-Predicate relationship characterizes the specific criteria of RDF Graph relationship, where two RDF Graphs $T_1$ and $T_2$ share common subject and predicate. The significance of these criteria is that two statements are semantically equivalent from the Subject and its property perspective. The only difference exists in a point that the two statements have different values for the same properties of the same subjects. It is evident that this criterion dictates a strong relationship between two RDF Graphs $T_1$ and $T_2$ and between two corresponding statements as well.

Conditions for Subject-Subject and Predicate-Predicate relationship are presented in Fig.1. in Venn diagram schema. Mathematically, there exists a Subject-Subject and Predicate-Predicate relationship between two RDF Graphs $T_1$ and $T_2$, if the following set theoretical expressions are all true.

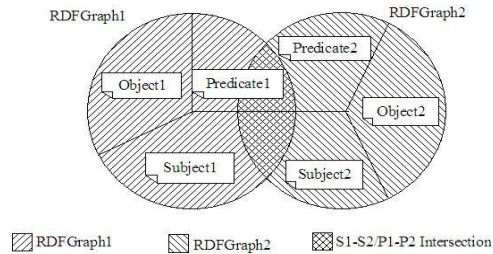

Fig. 1. Subject-Subject/Predicate-Predicate Relationship of two given RDF Graphs

$Sub(T_1) \cap Sub(T_2) \neq \varnothing$;
$Obj(T_1) \cap Obj(T_2) \neq \varnothing$;
$Sub(T_1) \cap Obj(T_2) \neq \varnothing$;
$Obj(T_1) \cap Sub(T_2) \neq \varnothing$;
$E_1 \cap E_2 \neq \varnothing$;

Following is an example of above mentioned relationship:

T1: Subject:http://www.example.org/staffid/85740
Predicate:http://www.example.org/terms/desig
Object:http://www.example.org/dept/accountant
T2: Subject:http://www.example.org/staffid/85740

Predicate:http://www.example.org/terms/desig
Object:http://www.example.org/club/treasurer

## 3.2 Object-Object and Predicate-Predicate relationship

Object-Object and Predicate-Predicate relationship identifies the criteria of RDF Graph relationship, where two RDF Graphs $T_1$ and $T_2$ share common object and predicate. The significance of this criterion can be exhibited in those cases where two statements are semantically equivalent from the property and its value perspectives. Two RDF Graphs related with this kind of condition, must have different subjects which hold same property with same values. Conditions for Object-Object and Predicate-Predicate relationships are presented in Fig.2. using Venn diagram schema.

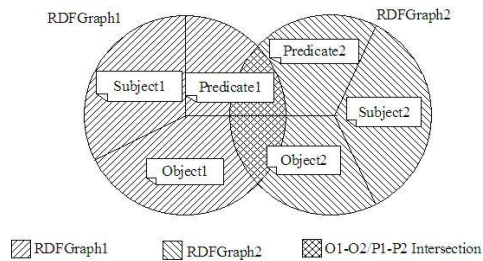

Fig. 2. Object-Object/Predicate-Predicate Relationship of two given RDF Graphs

Mathematically, there exists a Object-Object and Predicate-Predicate relationship between two RDF Graphs $T_1$ and $T_2$, if the following set theoretical expressions are all true.

$Obj(T_1) \cap Obj(T_2) \neq \varnothing$;
$Sub(T_1) \cap Sub(T_2) \neq \varnothing$;
$Sub(T_1) \cap Obj(T_2) \neq \varnothing$;
$Obj(T_1) \cap Sub(T_2) \neq \varnothing$;
$E_1 \cap E_2 \neq \varnothing$;

Following is an example of above mentioned relationship:

T1: Subject:http://www.example.org/staffid/85740
   Predicate: "published"
   Object:http://www.wikipedia.com/technology/C.V.
T2: Subject:http://www.example.org/staffid/85742
   Predicate: "published"
   Object:http://www.wikipedia.com/technology/C.V.

### 3.3 Subject-Predicate relationship

Subject-Predicate relationship has a different significance and consequence than the other two types of relationships discussed above. In this case, two RDF Graphs $T_1$ and $T_2$ never share their subject, object or predicate, rather the resource described by one's subject is same as that of resource described as predicate of others. With this condition, the subject of one statement acts as a property of the other statement. The two RDF Graphs related with their Subject - Predicate relation can represent complex indirect search construct. The two statements with completely different subjects could be linked with each other through this relationship.

Conditions for Subject-Predicate relationship are presented in Fig.3. In Venn diagram schema.

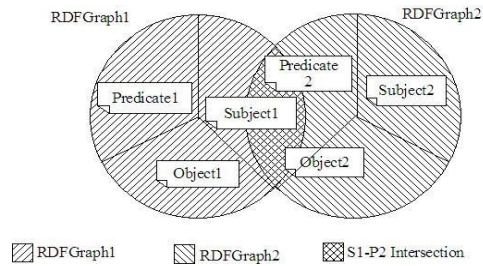

Fig. 3. Subject-Predicate Relationship of two given RDF Graphs

Mathematically, there exists a Subject-Predicate relationship between two RDF Graphs $T_1$ and $T_2$, if the following set theoretical expressions are all true.

$Sub(T_1) \cap Sub(T_2) \neq \varnothing$;
$Obj(T_1) \cap Obj(T_2) \neq \varnothing$;
$Sub(T_1) \cap E_2 \neq \varnothing$;
$E_1 \cap E_2 \neq \varnothing$;

Following is an example of above mentioned relationship:

T1: Subject:http://www.example.org/staffid/85740
    Predicate:http://www.example.org/terms/desig
    Object:http://www.example.org/dept/accountant
T2: Subject:http://www.example.org/terms/desig
    Predicate:http://www.example.org/staffid/85740
    Object:http://www.example.org/club/treasurer

### 3.4 Blank Node Relationship

When there is any anonymous resource presented in the triple, that is represented by Blank Node. Blank nodes are treated as simply indicating the existence of a thing, without using, or saying anything about, the name of that thing [16]. The blank nodes are reificated by splitting them in more than one different triple. Among these triples, the one which has the direct relationship with any other normal triple is called as Primary Triple and the rest of the triples which hold the transitive relation with that normal triple are called as Auxiliary triples. The link between these reificated triples with any other triple in RDF will be established by two relation scopes:

Relation between other triple and blank node primary triple: These are basically the normal N-Triple relationships which are already discussed in previous subsections (ie. SS-PP, OO-PP, SPand OP). Relation between reificated triples of blank node: These will be expressed by subject-object relationship.

Following is an example of RDF Graph relationship where a blank node is involved.

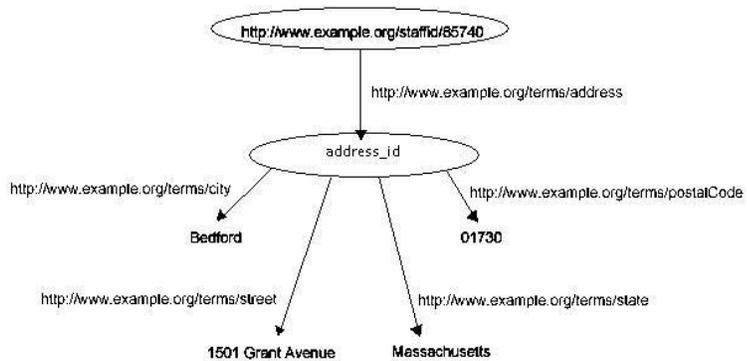

Fig. 4. Blank node reification for given RDF Graph

T1: Subject:http://www.example.org/staffid/85740
    Predicate: http://www.example.org/terms/address
    Object: address id
T2: Subject:http://www.example.org/staffid/85740
    Predicate:http://www.example.org/terms/desig
    Object:http://www.example.org/dept/accountant

In this example, in T1 Object: address-id is an anonymous resource or blank node and T2 is another normal triple. Now the T1 can be divided into a primary triple and a set of auxiliary triples like:

T1: Primary Relation: exsta :85740   exterms:address    :address id.

Set of Auxiliary Relations:
:address id      exterms:street      "1501 Grant Avenue" .
:address id      exterms:city        "Bedford" .
:address-id      exterms:state       "Massachusetts" .
:address-id      exterms:postalCode     "01730" .

From Fig.4. it is clear that, all the reificated triples are related with subject-object relationship and the primary reificated triple exsta :85740 exterms:address – :address id. is related with Triple2 by subject-subject relationship. In this context it can be concluded that, Triple2 is related with Triple1 by subject-subject/subject-object relationship. The example shown in Fig.4 illustrates that T1 and T2 are related with subject-subject/subject-object relationship as mentioned; this is only true for the current example. Generically the nature of the relationship will be any of the previously mentioned one or subject-object relation.

All the above set theoretic relations are based on subject-predicate-object and blank node relationship. This 1-Dimensional relation will be upgraded to n-Dimension in the next section implementing RDF Schema over the framework.

## 4 n-Dimensional relationship between RDF Graphs: Introducing RDF Schemas

In the previous section 1-D relational schema has been reported depending on mainly subject-predicate-object relationship. This relationship can be more re-ned to multidimensional status introducing RDF Property schema.

All four triples are mapped into RDF schema, which will be reported by RDF Graph. But this graph is not well formed because the system cannot guess that a Tiger is an animal. By introducing a new rdf:subclass of schema, a tertiary relationship has been induced in the graph. When there is no relationship between subject, object or predicate of n number of given triples, by applying these RDF Schemas over those triples, n-dimensional relation can be established among them. For an example the following natural language triples and their corresponding RDF Schematic representations can be considered:

Lion is an animal
Tiger is a cat
Zoos exhibit animals
Zoo1 exhibits the Tiger

| ex:Lion | rdf:type | ex:animal | (1) |
| ex:Tiger | rdf:type | ex:cat | (2) |
| zoo:exhibit | rdfs:range | ex:animal | (3) |
| ex:zoo1 | zoo:exhibit | ex:Tiger | (4) |

In Fig.5 by implementing RDF Schema, a new rdf:subclass of relation has been introduced in between triple(1) and triple(2) which was not in the actual RDF.

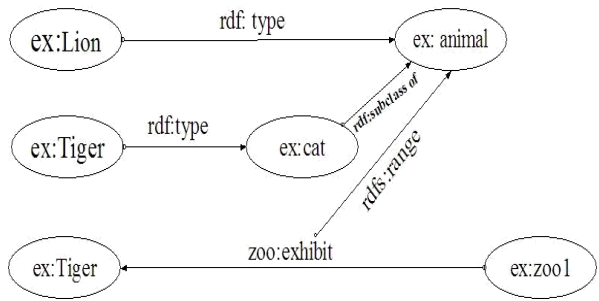

Fig. 5. Complete n-dimensional relation in RDF Graph

## 5 Conclusion

The current study has successfully met its objective of basic characterization of inter-relationship between two RDF Graphs. Set theory expressions have been identified which could be considered as necessary and sufficient conditions for discovering relationships between a RDF Graph pair. By introducing RDF Schema, the 1-dimensional relation can be upgraded to multi-dimensional induced relationships. The potential of this investigation can further be exploited with future research focusing on probabilistic quantification of these relationships and graph traversal based search algorithms within the network of such inter-related RDF Graphs.